%% file: ms.tex
\documentclass[aps,prd,twocolumn,showpacs,groupedaddress,nofootinbib]{revtex4-1}

\usepackage{graphicx}
\usepackage{dcolumn}
\usepackage{bm}
\usepackage{url}
\input{ads.tex}

\begin{document}

\title{Upper-twin-peak quasiperiodic oscillation in x-ray binaries and the energy from tidal circularization of relativistic orbits}

\author{C. German\`a}
\email{claudio.germana@gmail.com}
\email{claudio.germana@ufma.br}
\affiliation{Departamento de F\'isica, Universidade Federal do Maranh\~ao, 
S\~ao Lu\'is, MA, Brazil}

\date{\today}

\begin{abstract}
High frequency quasiperiodic oscillations (HF QPOs) detected in the power spectra
 of low mass x-ray binaries (LMXBs) could unveil the fingerprints of gravitation in
 strong field regime. Using the energy-momentum relation we calculate
 the energy a clump of plasma orbiting in the accretion disk releases during      
 circularization of its slightly eccentric relativistic orbit. Following previous works,
 we highlight the strong tidal force as mechanism to dissipate such
 energy. We show that tides acting on the clump are able to reproduce
 the observed coherence of the upper HF QPO seen in LMXBs
 with a neutron star (NS). The quantity of energy 
 released by the clump and relativistic boosting might give a 
 modulation amplitude in agreement with that observed 
 in the upper HF QPO. Both the amplitude and coherence of the upper
 HF QPO in NS LMXBs could allow us to disclose,
 for the first time, the tidal circularization of relativistic orbits 
 occurring around a neutron star.

\end{abstract}


\maketitle

\section{Introduction}\label{sec1}
The twin-peak high frequency quasiperiodic oscillations (HF QPOs), observed in the power spectra of
 low mass x-ray binaries (LMXBs) with either a neutron star (NS) or a black hole (BH),
 could carry information on the matter orbiting in the accretion disk around the compact object.
 Their central frequencies are typical of the orbital motion close to the compact object.
 HF QPOs are potential probes to prove the laws of gravitation close to a NS or a BH 
 \cite{2003ASPC..308..221L}.
 The first-discovered twin-peak HF QPOs were observed
 with central frequency up to $\sim1130$ Hz in a NS LMXB \cite{1996ApJ...469L...1V}. 
 They were named twin-peak kilohertz QPOs because they often show up in pairs. 
 The HF QPOs observed in BH LMXBs
 have frequencies of hundreds
 of hertz \cite{2006ARA&A..44...49R,2016ASSL..440...61B} and   
 show different features than HF QPOs seen in NS LMXBs. While in NS LMXBs the central
 frequency of the peaks is seen to vary, in BH LMXBs the twin-peak HF QPOs
 are observed at fixed frequencies, showing a cluster at the 3:2 frequency ratio. 
 The clustering has motivated models proposing that HF QPOs might be related 
 to resonance mechanisms of the matter orbiting in the curved space-time 
 \cite{2001A&A...374L..19A,2004ApJ...603L..89K,2005AN....326..830R,2006A&A...451..377H,2013A&A...552A..10S}. 
 The HF QPOs in BH LMXBs have a coherence lower than in NS LMXBs and an
 amplitude not displaying the characteristic patterns seen in NS LMXBs 
 (e.g. Refs.~\cite{2004astro.ph.10551V,2017MNRAS.468.2311M}).\\ 
 Low frequency QPOs ($< 100$ Hz) seen in both NS
 and BH LMXBs may be related to relativistic frame dragging
  around the spinning compact object
 \cite{1998ApJ...492L..59S}, a prediction of general relativity (GR) in strong field.
 The effect on the orbiting matter is known as Lense-Thirring precession \cite{1918PhyZ...19..156L}. 
 Recent works have put forward strong evidence that the low frequency QPO seen in the
 BH LMXB H1743-322 is produced by frame dragging \cite{2016MNRAS.461.1967I,2017MNRAS.464.2979I}. 
 In the case of NS LMXBs, recent data analysis shows that the predictions of
 the modeling differ from the data because other factors may affect the modulation
 mechanism \cite{2017MNRAS.465.3581V}.

 Other GR effects potentially detectable around the compact object in LMXBs are, 
 e.g., the periastron precession of the orbits \cite{1916AbhKP1916..189S}
 occurring on milliseconds time-scale as well as the existence of an
 innermost stable bound orbit (ISBO) \cite{misner,1990ApJ...358..538K}. The
 unprecedented opportunity to disclose such phenomena in the imprints left by the HF QPOs 
 has stimulated several works on the modulations that would be produced by matter 
 orbiting around a compact object 
 \cite{1998ApJ...508..791M,1999ApJ...524L..63S,2004ApJ...606.1098S,2014MNRAS.439.1933B}.
 Ray-tracing of the photons emitted by  
 an overbright hot-spot orbiting a Kerr black hole 
 shows the signal that a distant observer would see \cite{2004ApJ...606.1098S}. 
 The light curve produced by  the hot-spot is modulated at its orbital period 
 because of relativistic effects.
 Increasing the inclination towards 
 an edge-on view, the light curve becomes sharper because of increasing
 Doppler boosting and gravitational lensing.  
 The power spectrum of the signal 
 from a slightly eccentric orbit ($e\sim 0.1$) shows several
 peaks: the keplerian frequency $\nu_{k}$, the radial frequency\footnote{The radial 
 frequency $\nu_{r}$ is the 
 frequency of the cycle from periastron of the orbit to 
 apastron and back to periastron. In a curved space-time $\nu_{r}\neq\nu_{k}$.} 
 $\nu_{r}$, the beats $\nu_{k}\pm\nu_{r}$ and their harmonics. Also, the authors have 
 simulated the signal emitted by an arc sheared along the orbit.
 The power spectrum shows pronounced peaks at $\nu_{k}$ and $\nu_{k}\pm\nu_{r}$
 and much less power at the harmonics.\\ 
Ray-tracing presented in
 Ref.~\cite{2014MNRAS.439.1933B} shows the different detectability 
 that HF QPOs would have between current and future x-ray satellites,
 taking into account also the radial drift of the accreting hot-spot. In the power spectra 
 the peaks and harmonics at $\nu_{r}$, $\nu_{k}$ and $\nu_{k}+\nu_{r}$ (or $2(\nu_{k}-\nu_{r})$)
 are detected. Differences between the signal from the orbiting hot-spot and the 
 one from axisymmetrics disk oscillations are investigated as well.\\ 
 In a more dynamical framework, in Refs.~\cite{2005PhRvD..72j4024C,2009A&A...496..307K}
 were introduced ray-tracing techniques in the case of clumps of matter 
 stretched by the strong tidal force around a Schwarzschild black hole. 
 Differently than the rigid hot-spot case, the stretching of the clump, as long as it orbits,  
 leads to a sudden increase of its luminosity producing 
 a power law in the power spectrum \cite{2009A&A...496..307K,2009AIPC.1126..367G}. 
 Moreover, the stretching blurs the signal emitted in the case of a rigid sphere
 or a circular hot-spot. This implies some peaks and harmonics not be detected in the
 power spectrum. In other works the stretching of the clump is simulated as an arc along
 the orbit, in order to get power spectra with few peaks as
 in the observations. In the tidal model the stretching is a natural consequence of
 tidal deformation of the clump. The simulation using a slightly eccentric orbit
 ($e=0.1$) gives a power spectrum with a power law and two peaks,  
 as in the observations \cite{2006ARA&A..44...49R}. The lower peak in frequency 
 corresponds to $\nu_{k}$, the upper one to the beat $\nu_{k}+\nu_{r}$ \cite{2009AIPC.1126..367G}. 
 Tidal disruption events have already been recognized in the case of 
 stars disrupted by supermassive black holes 
 (e.g. Refs.~\cite{2014ApJ...783...23G,2015Natur.526..542M,2015JHEAp...7..148K,2017MNRAS.468..783L}). 
 Efforts to model such events are going 
 forward in the details (e.g. Refs.~\cite{2005ApJ...625..278G,2015ApJ...812L..39D,2017MNRAS.464.2816B,
 2017MNRAS.469.4879B}). QPOs have been detected in the energy 
 flux of some tidal disruption events \cite{2012Sci...337..949R,2014A&A...570L...2B}. 
 Tidal interaction is a mechanism that can provide significant amounts of energy. 
 In our neighborhood, some moons 
 display geological activities whose energy is pumped by the tidal force of the parent
 planet: the strongest volcanism in Jupiter's moon Io \cite{1979Sci...203..892P}
 and possibly the discovered ocean \cite{2008Icar..194..675R,2014Sci...344...78I} 
 and geothermal activity in 
 Saturn's moon Enceladus \cite{2006Sci...311.1422H,2014Icar..235...75S}.
 Thus, the strong tidal force by the compact object in LMXBs, acting on clumps
 of plasma orbiting in the accretion disk, may be a valid ingredient to  
 model the main features observed in twin-peak HF QPOs.

A planet/moon orbiting the central object on an eccentric orbit dissipates
 its orbital energy because of tides and its orbit gets circular
 \cite{2011MNRAS.415.2349R}. In Ref.~\cite{2008A&A...487..527C} has been
 shown that the orbit of a low-mass satellite around a Schwarzschild black hole
 circularizes and shrinks because of tides. Energy is transfered from 
 orbit to internal energy of the satellite. The energy emission 
 mechanism that would turn the released orbital energy into 
 electromagnetic radiation has been investigated 
 in Refs.~\cite{2009A&A...496..307K,2010AIPC.1205...30C}. The authors show that it may 
 be x-ray radiation from synchrotron mechanisms if the clump of plasma is permeated by
 a magnetic field. In Ref.~\cite{1989Ap&SS.158..205H} the authors conclude that
 magnetically confined massive clumps of plasma might form 
 in the inner part of the accretion disk. In Ref.~\cite{2013MNRAS.434..574S} it is shown 
 that the hard x-ray radiation, over 10-100 milliseconds time intervals,
 observed in two x-ray binaries is better interpreted through cyclo-synchrotron self-Compton mechanisms.
The calculations in Refs.~\cite{2009A&A...496..307K,2010AIPC.1205...30C} show that during 
 tidal stretching the magnetic field could largely increase. 
 Moreover, gravitational energy extracted through tides might go into 
 kinetic energy of the electrons in the plasma, since the clump is rapidly
 expanding into a pole. This mechanism could provide relativistic electrons emitting 
 synchrotron radiation. Magnetohydrodynamics simulations are required
 to know how this mechanism actually works. Recent numerical simulations of 
 the magnetic field in a star disrupted by tides 
 \cite{2017MNRAS.469.4879B} show a magnetic field largely increasing, 
 as from the calculations in Refs.~\cite{2009A&A...496..307K,2010AIPC.1205...30C}.
 
The emission of radiation because of the orbital energy released during tidal 
 circularization of the orbit thus would cause an overbrightness of the clump with respect
 to the background radiation from the disk.
In Ref.~\cite{2013MNRAS.430L...1G} has been shown that the timing law of the
 azimuth phase $\phi(t)$ on an slightly eccentric relativistic orbit produces multiple
 peaks in the power spectrum: the keplerian frequency $\nu_{k}$ and the
 beats $\nu_{k}\pm\nu_{r}$. The beats $\nu_{k}\pm\nu_{r}$ are produced because 
 of the eccentricity of the orbit. The orbiting body has a different orbital speed  
 at periastron and apastron passage, happening at the frequency $\nu_{r}\neq \nu_{k}$
 in the curved space-time. This introduces a modulation in the phase $\phi(t)$ at the
 relativistic radial frequency $\nu_{r}$. In the case of a circular  
 orbit (in every case in a flat space-time) only the peak at $\nu_{k}$  
 is produced. As already mentioned above, the timing law 
 $\phi(t)$ turns into a modulated observable light curve
 because of relativistic effects on the emitted photons 
 \cite{2004ApJ...606.1098S,2009A&A...496..307K,2014MNRAS.439.1933B}.
The amplitude of 
 the beats $\nu_{k}\pm\nu_{r}$ thus originates because of the orbital energy released 
 during tidal circularization of the orbit. 
 Moreover, the coherence of the beats is related to the
 time-scale the circularization takes place, 
 since once the orbit is circular or quasi-circular the beats 
 $\nu_{k}\pm\nu_{r}$ fade and the emitted energy is modulated only at the 
 keplerian frequency $\nu_{k}$.

Most efforts to interpret the twin-peak HF QPOs have focused on the identification 
 of their central frequencies with those of the orbital modes
 in the curved space-time. The proposed models link the upper HF QPO of the 
 twin-peaks to the keplerian modulation $\nu_{k}$ produced by a clump of plasma orbiting 
 in the accretion disk,
 other models link the lower HF QPO to $\nu_{k}$ 
\cite{1998ApJ...508..791M,1999ApJ...524L..63S,2004ApJ...606.1098S,1999ApJ...522L.113O,1999ApJ...518L..95T,2003ApJ...584L..83M}.
 On the other hand, attempting to interpret the amplitude and coherence of HF QPOs 
 might disclose useful information on their nature as well.
In Refs.~\cite{2001ApJ...561.1016M,2006MNRAS.370.1140B,2006MNRAS.371.1925M,2011ApJ...728....9B}  
 are reported both the amplitude and coherence of the twin-peak HF QPOs observed   
 in NS LMXBs. The behavior of the amplitude as a function of the central
 frequency of the peaks shows characteristic patterns
 in atoll NS LMXBs \cite{1989A&A...225...79H}.
 The amplitude of the upper HF QPO displays
 a decrease with increasing central frequency of the peak,
 instead the amplitude of the lower HF QPO shows an increase and then a decrease.  
The coherence $Q$ of the lower HF QPO  
 ($Q=\nu/\Delta\nu$ with $\nu$ central frequency and $\Delta\nu$ full width at
 half maximum of the peak) shows a characteristic pattern too: 
 $Q$ as a function of $\nu$ increases
 and then drops abruptly \cite{2006MNRAS.370.1140B,2011ApJ...728....9B,2006MNRAS.371.1925M}. 
 In Ref.~\cite{2006MNRAS.370.1140B} has been underlined that the abrupt drop 
 of $Q$, seen in several atoll NS LMXBs, could be a signature 
 of the oscillation approaching the ISBO predicted by GR. 
 This relevant issue was subsequently discussed with extensively data analysis 
 in Refs.~\cite{2006MNRAS.371.1925M,2011ApJ...728....9B}.
 Although the excursion of $Q$ of the lower HF QPO 
 is more than an order of magnitude, the $Q$ of
 the upper HF QPO shows an almost flat trend over a large range of frequencies,
 mostly remaining of the order of $Q\sim10$.\\
 In a previous work (Ref.~\cite{2015PhRvD..91h3013G}, hereafter GC15)
 we have proposed that the amplitude and coherence of the lower HF QPO might 
 originate from the energy released by a clump of plasma spiraling to inner 
 orbits because of the work done by the tidal force, dissipating the orbital energy. 
 In this paper we aim to investigate on the amplitude and coherence of the upper HF QPO
 \cite{2006MNRAS.370.1140B}. Here is proposed that the upper HF QPO might originate 
 from the energy released during tidal circularization of the clump's orbit.
 In Ref.~\cite{2008A&A...487..527C} has been shown that the orbit of a clump of matter  
 orbiting a Schwarzschild black hole
 circularizes and shrinks because of tides. 
 The release of orbital energy during circularization of the orbit 
 might provide the overbrightness of the clump required in order to 
 produce detectable modulations \cite{2004ApJ...606.1098S,2009A&A...496..307K}.   
The emitted photons are modulated at $\nu_{k}$ and $\nu_{k}\pm\nu_{r}$ in the power spectrum 
 \cite{2004ApJ...606.1098S,2013MNRAS.430L...1G}. 
 The beats $\nu_{k}\pm\nu_{r}$ should show up 
 only in the phase of tidal circularization of the orbit, since once the 
 orbit gets circular $\nu_{k}\pm\nu_{r}$ fade and the emitted radiation is modulated 
 only at $\nu_{k}$. Tidal disruption simulations show an  
 upper HF QPO corresponding to the beat $\nu_{k}+\nu_{r}$ \cite{2009AIPC.1126..367G}.
Therefore, we believe and inspect that 
 both the amplitude and coherence of the upper HF QPO 
 in the observations \cite{2006MNRAS.370.1140B} should be 
 reproduced by the energy released during tidal circularization 
 of relativistic orbits.\\
 The paper is organized as follows. In Section~\ref{sec2}
 we recall the main arguments described in GC15
 about the tidal load on clumps of plasma orbiting in the accretion disk.
 In Section~\ref{sec3} we explore the idea presented in this manuscript,
 i.e. the amplitude and coherence of the upper HF QPO seen in NS LMXBs 
 could be related to the energy released during 
 tidal circularization of relativistic orbits. We estimate the energy
 released by an orbiting clump of plasma when its slightly eccentric orbit gets  
 circular.
We use the energy-momentum relation in the Schwarzschild metric\footnote{We 
 use the Schwarzschild metric because there are in the
 literature exact parametrization of both the orbital energy $E$ and angular
 momentum $L$ per unit mass, for a test-particle on an orbit with generic eccentricity $e$
 \cite{1994PhRvD..50.3816C}.}
 since it is the relativistic equation that embeds all the contributions to the total energy 
 of an orbiting clump of matter.  
 The time-scale of tidal circularization of the orbit is calculated. 
 Afterwards, we calculate the coherence $Q$ the produced beat $\nu_{k}+\nu_{r}$ would have.
 We compare it to the upper HF QPO coherence pattern seen in the observations 
 (e.g. Fig.~2 in Ref.~\cite{2006MNRAS.370.1140B}).
 In Section~\ref{sec4} we attempt to tie the orbital energy released\footnote{We emphasize 
 that the main goal of the manuscript is to justify how (from where) the amount
 of energy carried by the detected
 upper HF QPO would originate. We estimate the amount 
 of the bolometric energy that would be released by this mechanism, to compare it to the 
 bolometric energy observed in the upper HF QPO. Here our main purpose is not the spectral energy distribution 
 (i.e. how the orbital energy then is emitted), 
 which depends on the exact energy emission mechanism
 (see Sec.~IV for a discussion on this point).} during 
 circularization of the orbit to the observable fraction of energy modulated by Doppler
 boosting. We follow 
 the detailed results in Ref.~\cite{2004ApJ...606.1098S} to get the observable
 amplitude of the beat $\nu_{k}+\nu_{r}$.  
In Section~\ref{sec5} we discuss the results in this paper 
 in light of other theoretical and observational results.
 Section~\ref{sec6} summarizes the conclusions.
        
\section{Orbiting clumps of plasma and tidal load}\label{sec2}

 Motivated by the results from tidal disruption of clumps orbiting a  
 Schwarzschild black hole \cite{2005PhRvD..72j4024C,2009A&A...496..307K},
 reproducing power spectra much alike to the 
 observed ones \cite{2009AIPC.1126..367G}, in GC15 we have estimated the
 energy coming from the tidal disruption of a 
 clump of plasma in the accretion disk around LMXBs.
 Magnetohydrodynamics simulations show that the inner part of the accretion disk is highly turbulent
 \cite{2001ApJ...548..348H}. In Ref.~\cite{2013Sci...339.1048C} the authors reported the discover
 of large structures in the accretion disk of a x-ray binary.
 Propagating accretion rate fluctuations in the disk are modeled
 \cite{2013MNRAS.434.1476I,2016AN....337..385I} 
 to reproduce the aperiodic variability observed in BH LMXBs.  
 Thus, it is hard thinking to a
 smooth accretion disk, but rather it may be characterized by inhomogeneities propagating throughout it.
 Note that in Ref.~\cite{1989Ap&SS.158..205H} is shown that magnetically confined
 massive clumps of plasma might form in the inner part of the accretion disk.  
 In light of this, in GC15 we explored the idea of treating a clump of plasma as characterized
 by some internal force keeping the clump together 
 (e.g. electrochemical bounds and/or magnetic forces). In this section we recall 
 the main arguments in GC15.
 
 A spherical clump of radius $R$, mass $\mu$ and density $\rho$ undergoes a tidal force  
 (between two opposite spherical caps of the clump, at $r-R$ and $r+R$; see also GC15)
\begin{eqnarray}\label{eq1}
F_{T}&=&\mu'c^{2}\left[\left(\frac{dV_{eff}}{dr}\right)_{(r-R)}-\left(\frac{dV_{eff}}{dr}\right)_{(r+R)}\right]\\ \nonumber
&\approx&\mu'c^{2}2R\left(\frac{d^{2}V_{eff}}{dr^{2}}\right)_{r}
\end{eqnarray} 
 where $\mu'=\rho V'$ is the mass of the spherical cap, of height, say, one tenth of the radius, $h=R/10$. 
 The volume of the cap is $V'= \pi h^{2}(R-h/3)$. $V_{eff}$ in (\ref{eq1}) is the 
 gravitational effective potential in the Schwarzschild metric (\ref{eq7}). In the case of  
 a solid-state clump of matter, the clump is kept together by an internal force (electrochemical bounds) 
 characterized by the ultimate tensile strength $\sigma$, i.e. 
 the internal force per unit area. The tidal force has to be weaker than 
 internal forces, $F_{T}\leq2\pi R h\sigma$. From 
 this inequality we can get some order of magnitude on the maximum radius $R_{max}$ set by tides
 \footnote{Note that the $R_{max}$ calculated  
 in GC15 in the case of a solid-state clump agrees to the dimensions derived 
 in Ref.~\cite{2009NCimB.124..155K} of a bar falling into a gravitational field.}
\begin{eqnarray}\label{eq2}
R_{max}&=&\left(10\left(1-\frac{1}{30}\right)^{-1}\frac{c_{s}^{2}}{c^{2}}\frac{\sigma}{Y}\times\right.\\ \nonumber
&&\left.\left(-\frac{2m}{r^{3}}+\frac{3\tilde{L}^{2}}{r^{4}}-\frac{12m\tilde{L}^{2}}{r^{5}}\right)^{-1}\right)^{1/2}
\end{eqnarray}
 where we wrote the density $\rho=c_{s}^2/Y$, $Y$ is the Young's modulus of the material, $c_{s}$ the 
 speed of sound in it. As mentioned above, in Section IV of GC15 we explored the idea of treating clumps of plasma 
 in the accretion disk as characterized by some internal force per unit area $\sigma$ 
 (electrochemical bounds and/or magnetic forces). The speed of sound in the plasma is \cite{2002apa..book.....F} 
\begin{equation}\label{eq3}
c_{s}=\left(\frac{\gamma Z k T}{m_{i}}\right)^{1/2}    
\end{equation}
\begin{figure}[!t!]
\includegraphics[width=0.47\textwidth]{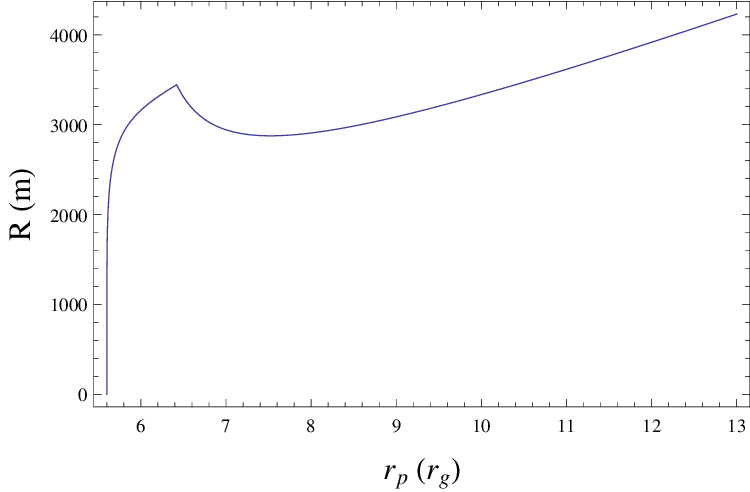}
\caption{Radius $R$ set by tides of a clump of plasma as a 
 function of the periastron of the orbit $r_{p}$, around a $2\ M_{\odot}$ neutron star.}\label{fig1}
\end{figure}
 where $\gamma\sim5/3$ is the adiabatic index, $Z$ the charge state ($Z=1$ for
 a hot plasma), $m_{i}$ the ion hydrogen mass, $k$ the Boltzmann's constant \cite{2002apa..book.....F}.\\
 In CG15 we pointed out that clumps with $R=R_{max}$ would not probably form at all 
 because of tides. The tidal load (the tidal force (\ref{eq1}) per unit area) has to be $n$ times 
 smaller than $\sigma$, i.e. $F_{T}/2\pi R h=\sigma_{T}=\sigma/n$, where
\begin{eqnarray}\label{eq4}
\sigma_{T}&=&\frac{\mu'c^{2}}{2\pi Rh}\left[\left(\frac{dV_{eff}}{dr}\right)_{(r-R)}-\left(\frac{dV_{eff}}{dr}\right)_{(r+R)}\right]\\ \nonumber
&\approx&\frac{10\mu'c^{2}}{\pi R}\left(\frac{d^{2}V_{eff}}{dr^{2}}\right)_{r}
\end{eqnarray}
 In GC15 we constrained $n=5$ as upper limit, giving $R\sim3000$ m. A larger 
 $n$ implies a clump with radius $R$ emitting gravitational energy 
 lower than that observed in HF QPOs ($\approx10^{35}-10^{36}$ erg/s).
 On the other hand, a smaller $n$ gives larger radii $R$, close to $R_{max}$.
 As mentioned above, such clumps would not probably form/survive at all because of tides.  
 Fig.~\ref{fig1} shows the radius $R=R_{max}/\sqrt 5$ set by tides
 (from equation (\ref{eq2})) as a function of the periastron $r_{p}$ of the orbit,
 in the case $\sigma_{T}=\sigma/5$. In (\ref{eq2})
 the ratio $\sigma/Y$ was constrained in GC15 (equation (9)) and is  
 $\sigma/Y=300$ in atoll sources ($\sigma/Y=70$ in Z-sources; see Section~VII B in GC15).
 The speed of sound $c_{s}$ is from (\ref{eq3}). In Fig.~\ref{fig1} we see that,
 as long as the tidal force strengthen towards the inner regions, 
 $R$ decreases as expected. However, getting closer to ISBO ($r\sim 5.6\ r_{g}$) $R$
 increases and then drops. The slightly increase is caused by the weakening of the 
 tidal force close to ISBO. Close to ISBO the gravitational potential
 (\ref{eq7}) flattens and, therefore, the tidal force weaken.
 This can be seen in Fig.~\ref{fig2}. It shows the tidal load            
 $\sigma_{T}$ (\ref{eq4}) in Pascal on a clump of plasma $R=3000$ m big for several orbits
 of different periastron. Over each orbit (each segment in the figure) 
 $\sigma_{T}$ changes from the periastron to the apastron of the orbit. Its overall
 behavior increases and then drops close to ISBO because of the flattening of the potential.
 The flattening of the minimum of the potential $V_{eff}$ is a features of GR
 \cite{misner} and causes the decrease of the difference 
 of potential energy between close orbits reported in GC15.\\ 
 The drop of $R$ in Fig.~\ref{fig1}
 close to ISBO is caused by the drop at ISBO (inner edge of the accretion disk) of 
 the speed of sound in the plasma (see equation (\ref{eq2})). 
 The cusp seen at $r_{p}\sim 6.4\ r_{g}$
 is because of the orbit at 
 which the tidal force is almost equal at periastron and apastron. Orbits 
 with bigger radii have the tidal force stronger at periastron,
 as expected, therefore we calculate the radius
 $R$ of the clump set by tides at periastron. 
 However, orbits with $r$ smaller than $r\sim 6.4\ r_{g}$ have  
 a tidal force stronger at apastron, because of the flattening of the potential.
 This can been see in Fig.~\ref{fig2}. Thus, we calculate the radius $R$ set by
 tides at the apastron of the orbit.\\
 The patterns in both figures are for orbits of eccentricity\footnote{It might be reasonable 
 thinking that during accretion a clump of plasma may have a trajectory on a not perfect circular orbit, 
 because of the turbulent environment \cite{2001ApJ...548..348H}.
 Numerical simulations in Ref.~\cite{2004ApJ...606.1098S} 
 reproduce multiple peaks in the power spectrum, 
 at $\nu_{k}$ and the beats $\nu_{k}\pm\nu_{r}$, for orbits with small eccentricity ($e\sim0.1$). 
 In Ref.~\cite{2009AIPC.1126..367G} the   
 upper HF QPO at $\nu_{k}+\nu_{r}$ in the power spectrum from numerical simulations
 is reproduced for an orbit with eccentricity $e=0.1$. Such results
 \cite{2004ApJ...606.1098S,2009AIPC.1126..367G}, much alike to observations,
 suggest that clumps on orbit with low $e$ may exist in the disk. 
 So here we chose $e=0.1$, also to pursue the results reported in GC15 and 
 Ref.~\cite{2013MNRAS.430L...1G}.} 
 $e=0.1$, for a neutron star of $2\ M_{\odot}$ and an accretion rate
 of $\dot{M}\sim7\times 10^{16}$ g/s, 
 giving the luminosity observed in atoll sources, i.e. $L\sim 0.07\ L_{Edd}\sim10^{37}$ erg/s
 (\cite{2010MNRAS.408..622S}, where $L_{Edd}\sim2.5\times10^{38}$ erg/s
 is the Eddington luminosity for a
 $\sim2\ M_{\odot}$ neutron star \cite{2002apa..book.....F}). This accretion rate gives  
 a density of the clump of plasma in the accretion disk of $\rho\sim 1$ g/cm$^{3}$
 and the speed of sound in it $c_{s}\sim4\times10^{7}$ cm/s \cite{2002apa..book.....F}. 
\begin{figure}[t]
\includegraphics[width=0.47\textwidth]{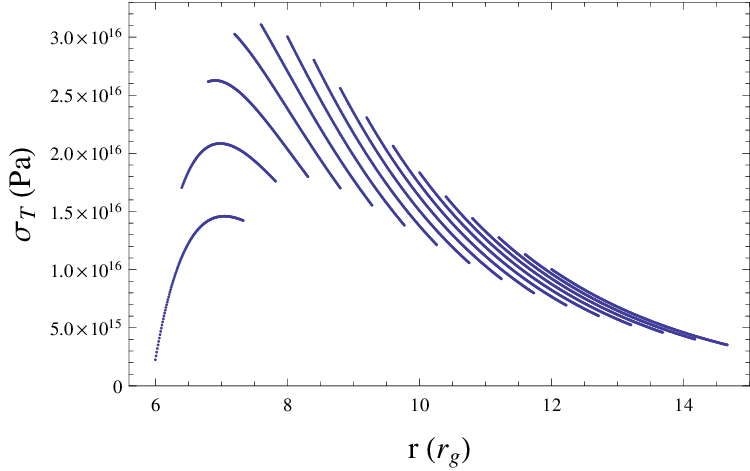}
\caption{Tidal load $\sigma_{T}$ (\ref{eq4}) over a clump of plasma $R=3000$ m big as a function of the orbital radius.
 Each segment draws the variation of the load from periastron to apastron of the orbit.}\label{fig2}
\end{figure}

\section{The energy and time-scale from tidal circularization of relativistic orbits}\label{sec3}

The total energy of an orbiting test-particle of mass $\mu$ is
 enclosed in the energy-momentum relation\footnote{Hereafter
 we use geometric units ($G=c=1$), unless differently specified.} 
\begin{equation}\label{eq5}
g_{\alpha\beta}p^{\alpha}p^{\beta}=-\mu^{2}
\end{equation}
with $g_{\alpha\beta}$ metric tensor and $p^{\alpha(\beta)}$ contravariant
 four-momentum of the test particle \cite{misner}. In the Schwarzschild
 metric substituting the $p^{\alpha}=dx^{\alpha}/d\tau$ 
 ($\tau$ proper time; see e.g. \cite{1994PhRvD..50.3816C}) and extending (\ref{eq5}) we get 
\begin{equation}\label{eq6}
\tilde{E}^{2}=\left(\frac{dr}{d\tau}\right)^{2}+\left(1-\frac{2m}{r}\right)\left(1+\frac{\tilde{L}^{2}}{r^{2}}\right)
\end{equation}
$m$ is the mass of the compact object\footnote{The mass $m$ in geometric units is equal to 
 the gravitational radius of the compact object $r_{g}=GM/c^{2}$, where $M$ 
 is the mass of the compact object in international system units, $G$ the
 gravitational constant and $c$ the speed of light. For a $2\ M_{\odot}$ 
 neutron star $r_{g}\sim3$ km.}, $\tilde{E}$ and $\tilde{L}$ total
 energy and angular momentum per unit mass $\mu$ of the test-particle, 
 $r$ is the radial coordinate.
 Equation (\ref{eq6}) (whose square root, multiplied by $\mu c^{2}$, we can
 write as $E=\mu_{rel}c^{2}$, with $\mu_{rel}$ relativistic mass) tells
 us the energy contributions to the total energy $\tilde{E}$.
 The first term is the energy coming from the radial motion,
 i.e. the motion from periastron to apastron and back to periastron.
 The second term is the effective gravitational potential \cite{1994PhRvD..50.3816C}
 \begin{equation}\label{eq7}
 V_{eff}=1-\frac{2m}{r}-\frac{2m\tilde{L}^{2}}{r^{3}}+\frac{\tilde{L}^{2}}{r^{2}}
 \end{equation}
 with contribution from the rest-mass energy (per unit mass $\mu$), the gravitational 
 and centrifugal potential.
 
In Ref.~\cite{1994PhRvD..50.3816C} are reported exact parametrization
 for the total (or orbital) specific energy $\tilde{E}$ and specific
 angular momentum $\tilde{L}$, for a generic orbit of semi-latus
 rectum $p$ and eccentricity $e$
\begin{equation}\label{eq8}
\tilde{E}\left(p,e\right)=\left(\frac{\left(p-2-2e\right)\left(p-2+2e\right)}{p\left(p-3-e^{2}\right)}\right)^{1/2}
\end{equation}

\begin{equation}\label{eq9}
\tilde{L}\left(m,p,e\right)=\left(\frac{p^{2}m^{2}}{p-3-e^{2}}\right)^{1/2}
\end{equation}
$p$ is linked to the periastron $r_{p}$ of the orbit through $r_{p}=pm/(1+e)$.\\
The energy (in international system units) that a clump of matter of 
 mass $\mu$ would release,
 if its orbit of eccentricity $e$ is circularized, is from (\ref{eq8})  
\begin{equation}\label{eq10}
\epsilon=\mu c^{2}\left(\tilde{E}\left(p,e\right)-\tilde{E}\left(p,0\right)\right)
\end{equation}  

We aim to compare the released energy $\epsilon$ to the energy (amplitude)
 carried by the upper HF QPO observed in NS LMXBs
 (Fig.~3 in Ref.~\cite{2006MNRAS.370.1140B}). The upper HF QPO
 of the twin-peaks corresponds to the beat $\nu_{k}+\nu_{r}$ in the power 
 spectrum from numerical simulations \cite{2009AIPC.1126..367G,2013MNRAS.430L...1G}. 
 This beat is caused by the eccentricity of the orbit and originates
 only in the phase of tidal circularization of the orbit,
 when energy is released and it is modulated at $\nu_{k}+\nu_{r}$,
 till the orbit gets circular, then $\nu_{k}+\nu_{r}$ fades.\\
We calculate the relativistic keplerian\footnote{We would warn that keplerian motion
 for matter orbiting close to a neutron star is an approximation, since the effects of a boundary layer might 
 deviate the orbital motion from purely keplerian.} $\nu_{k}$ and radial $\nu_{r}$ frequency 
\begin{figure}[!t!]
\includegraphics[width=0.47\textwidth]{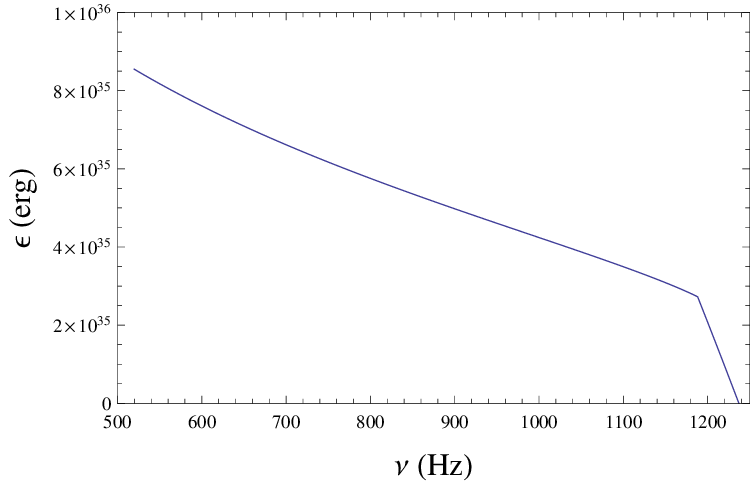}
\caption{Orbital energy released by a clump of plasma with $R$ as in Fig.~\ref{fig1} in order to circularize its orbit
 with initial $e=0.1$. 
 The energy is plotted as a function of the frequency of the beat $\nu_{k}+\nu_{r}$ for a $2\ M_{\odot}$ compact object.}\label{fig3}
\end{figure}
 as in Ref.~\cite{2013MNRAS.430L...1G}, for an orbit with eccentricity $e=0.1$. 
 Fig.~\ref{fig3}
 shows the orbital energy released $\epsilon$ (\ref{eq10}) to circularize the orbit of
 initial $e=0.1$ as a function of the frequency of the beat $\nu_{k}+\nu_{r}$, 
 i.e. for clumps orbiting at different orbital radii. 
 The range of orbital radii is $\sim$6 $r_{g}$ to 13 $r_{g}$. 
 At each orbital radius the clumps have $R$ as in Fig.~\ref{fig1}.
 The energy released corresponds to $\sim 0.3\%\ \mu c^{2}$. 
 We see that the energy released when, e.g., $\nu_{k}+\nu_{r}\sim 520$ Hz 
 ($r_{p}\sim 13\ r_{g}$) is higher than that released by a
 clump orbiting at $r_{p}\sim 7\ r_{g}$ ($\nu_{k}+\nu_{r}\sim 1100$ Hz). Close to ISBO it drops.\\ 
 With the amount of orbital energy released by the clump to circularize its orbit
 we can investigate whether the upper HF QPO seen in the observations 
 could actually originate from tidal circularization of relativistic orbits. 
 We calculate the time-scale the circularization 
 of the orbits by tides would take place. Then we compare the derived coherence
 of $\nu_{k}+\nu_{r}$ to the coherence behavior of the upper HF QPO observed
 in several atoll NS LMXBs (Fig.~2 in Ref.~\cite{2006MNRAS.370.1140B}).\\ 
 The tidal force removes energy from orbit and loads it on the 
 clump (e.g. Ref.~\cite{1977ApJ...213..183P,2008A&A...487..527C}).
 We aim to estimate the energy loaded by tides on the clump over one radial cycle, 
 from periastron to apastron and back to periastron (see Fig.\ref{fig2}). 
 To get order of magnitude, we substitute into (\ref{eq4})
 the parametrized radius $r(\chi)=pm/(1+e\cos(\chi))$ as a function of the radial phase $\chi$
 \cite{1994PhRvD..50.3816C}. The tidal load ($\ref{eq4}$) as a function of
 $\chi$, which is an energy per unit volume,
 is integrated over one radial cycle $\chi$, from periastron ($\chi=0$) to apastron
 ($\chi=\pi$) and back to periastron 
 ($\chi=2\pi$). We multiply for the volume of the clump to get the energy
 loaded by tides per periastron passage.
For a clump with $R$ as in Fig.~\ref{fig1} and a density of the plasma
 typical for an atoll source ($\rho\sim 1$ g/cm$^{3}$),
 the estimated amount of energy is of the order of \footnote{Note that the order 
 of magnitude obtained $E_{tide}\sim 0.1\%\ \mu c^{2}$ agrees to that calculated with other formalisms 
 in the case of a star disrupted by a supermassive black hole \cite{2005ApJ...625..278G}.} $E_{tide}\sim10^{35}$ erg. 
 We divide $\epsilon$ from (\ref{eq10}) by $E_{tide}$ (as a function of the orbital radius)  
 to get the number of periastron passages $N$ in order to circularize the orbit.
 The time it takes to circularize the orbit then is $t'=N/\nu_{r}$, equal to\footnote{This time-scale 
 is in agreement with that from the calculations in 
 Ref.~\cite{2008A&A...487..527C}.} 
 $t'\sim 0.01$ s at $r\sim 8\ r_{g}$.
 The coherence of the beat $\nu_{k}+\nu_{r}$ is 
 $Q=(\nu_{k}+\nu_{r})/\Delta\nu=(\nu_{k}+\nu_{r})t'$.  
 Fig.~\ref{fig4} shows the coherence $Q$ obtained from our calculations 
 as a function of the frequency $\nu_{k}+\nu_{r}$.
 Like in Fig.~\ref{fig3}, the range of frequency corresponds to a range of orbital radii of 
 $\sim 13-5.6\ r_{g}$. The radius $R$ of the clump is shown in Fig.~\ref{fig1}. 
 The coherence $Q$ is mostly constant and of the order of 10. 
 Both its value and
 trend are much alike to the coherence of the upper HF QPO observed in NS LMXBs, Fig.~2
 of Ref.~\cite{2006MNRAS.370.1140B} (filled star symbols). In Fig.~2 of Ref.~\cite{2006MNRAS.370.1140B}
 $Q$ is of the order of $Q\sim 10$ for most of the sources.
\begin{figure}[!tt!]
 \includegraphics[width=0.47\textwidth]{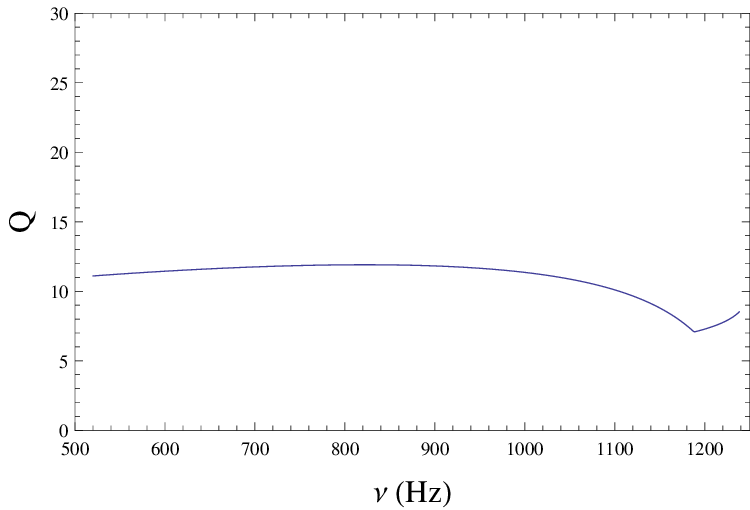}
 \caption{Simulated coherence $Q$ of the beat $\nu_{k}+\nu_{r}$ as a function of the 
 frequency $\nu_{k}+\nu_{r}$ for a $2\ M_{\odot}$ compact object. The value of $Q$ is
 related to the time scale the tidal circularization of relativistic orbits takes place, 
 for clumps of plasma of $R$ as in Fig.~\ref{fig1}.
 Such behavior is typical of the $Q$ of the detected upper HF QPO. For a comparison
 with the data see Fig.~2 in Ref.~\cite{2006MNRAS.370.1140B}}.\label{fig4}
\end{figure}

We see that the $Q$ calculated here strongly depends on the radius $R$ of the clump, $Q\propto R^{-2}$.
 It may be worth emphasizing that the $R$ in Fig~\ref{fig1} is derived from the calculations in Section~\ref{sec2} 
 and the way to derive it was described in Sections~III, IV in GC15. 
 We are not assuming 
 an $R$ to match the $Q$ from the observations, but its value is derived from calculations. This may be a 
 significant result within this framework. Indeed, from the calculated $R$ this approximated modeling is able to 
 give for the first time both $Q$ and the amplitude of the upper HF QPO (see Sec.~\ref{sec4})
 in agreement with those from observations \cite{2006MNRAS.370.1140B}.

\section{Tying the released orbital energy to the observable amplitude of the modulation}\label{sec4}

 The orbital energy released during
 tidal circularization of the orbit (eq.~\ref{eq10})
 gives time-scales of dissipation in agreement with the coherence $Q$ of the upper HF QPO
 detected in atoll NS LMXBs \cite{2006MNRAS.370.1140B}. 
 However, this released orbital energy has to be converted somehow to electromagnetic
 radiation in order the upper HF QPO to be detected. Moreover, only a fraction of this radiation 
 is modulated by Doppler boosting and detectable as HF QPOs \cite{2004ApJ...606.1098S}. 
 In this section we discuss how the extracted orbital energy by tides
 would turn into radiation emitted by the clump (see footnote 3). 
 We also estimate the amount of energy that would be modulated 
 by Doppler boosting and detected as a QPO, following the results in Ref.~\cite{2004ApJ...606.1098S}. 

 From the energy emission spectra of LMXBs we know that HF QPOs are observed in hard x-ray, 
 their amplitude keep increasing towards hard x-ray ($>$ 5 keV \cite{1996ApJ...469L..13B}). 
 This means that the only thermal emission from the disk (soft x-ray, $\sim 1$ keV) can 
 not justify their nature. A corona of hot electrons and/or a boundary 
 layer contribute to the energy emission spectra observed  
 in LMXBs (see e.g. Refs.~\cite{2001ApJ...547..355P,2013MNRAS.432.1144S}). 
 These components
 produce the hard x-ray spectrum seen in LMXBs. The soft x-ray photons from
 the disk are inverse-Compton scattered to higher energy by the corona and/or the boundary layer. 
 There are evidences that the same mechanism could amplify the amplitude of 
 the HF QPOs at hard x-ray \cite{2001ApJ...549L.229L,2005AN....326..812G}. 
 That is, the HF QPOs could be produced in the disk, but then they are
 amplified to hard x-ray by the corona and/or the boundary layer.
 It was recently suggested that the occurrence of the lower HF QPO could be 
 because of some resonance between the comptonising medium and the accretion disk and/or 
 the neutron star surface \cite{2017MNRAS.471.1208R}. 
 On the other hand, in Ref.~\cite{2005tsra.conf..511S} 
 it is shown, by means of Monte Carlo ray-tracing, that multiple scattering of soft photons 
 from the disk in a corona of hot electrons would smooth the oscillation that originates in the disk.  
 In Ref.~\cite{2005tsra.conf..511S} it is
 suggested that it is unlikely that the same mechanism would produce HF QPOs at hard x-ray, 
 since the emerging hard x-ray suffered more scattering than soft x-ray, thus the oscillation 
 has a low amplitude at high energy bands (see Fig.~5 in Ref.~\cite{2005tsra.conf..511S}). 
 It is also suggested that there may be in the disk a hot-spot already emitting hard x-ray photons,
 such that they are moderately scattered by the surrounding corona.    
 In Ref.~\cite{2013MNRAS.434..574S} the authors studied the energy spectra of two x-ray binaries over 
 $10-100$ ms time-scales. They concluded that the hard x-ray radiation is better 
 explained through cyclo-synchrotron self-Compton mechanisms. Thus, if clumps of plasma 
 in the accretion disk are permeated by some magnetic field, tidal stretching of the clump may 
 provide a mechanism to produce non-thermal electromagnetic radiation. 
 The orbital energy extracted through tides (e.g. Fig.~\ref{fig3}) is transferred into 
 internal energy of the clump (e.g. Refs.~\cite{1977ApJ...213..183P,2008A&A...487..527C}). 
 In Refs.~\cite{2009A&A...496..307K,2010AIPC.1205...30C} it is shown that during 
 tidal stretching the magnetic field could largely increase. 
 The extracted orbital energy could go into kinetic energy of the electrons in the plasma, 
 since the clump is rapidly expanding into a pole. 
 This mechanism could provide relativistic electrons winding around the 
 magnetic filed of the clump and producing synchrotron radiation \cite{2009A&A...496..307K,2010AIPC.1205...30C}. 
 Synchrotron radiation by compact hot-spots has already been proposed as
 mechanism to produce the hard x-ray spectrum 
 seen in HF QPOs \cite{2011RAA....11..631Y}. It is clear that 
 full magnetohydrodynamics simulations are required to see how the clump disrupted by tides 
 would emit its energy. On the other hand, we have some clues which could be used 
 to estimate the magnetic field the clump 
 would have and checking whether it is consistent with that measured in LMXBs ($B\sim10^{8}-10^{13}$ G \cite{1999A&AT...18..447P,2015SSRv..191..293R}).  
 In Section~IV of GC15 we explored the idea of treating the clump of plasma as characterized by some internal force 
 keeping it together, e.g. electrochemical bounds and/or a magnetic force. 
 In Ref.~\cite{1989Ap&SS.158..205H} is pointed out that magnetically confined massive clumps of plasma might 
 form in the inner part of the accretion disk. We calculated in equation (9) in GC15 
 the value of the ratio $\sigma/Y$, where $\sigma$ is the internal force per unit area, $Y=\rho c_{s}^{2}$
 is the Young's modulus of the material
\begin{eqnarray}\label{eq11}
\frac{\sigma}{Y}&=&\left(\frac{3}{4\pi}\left(\frac{1}{10}\right)^{3/2}\left(1-\frac{1}{30}\right)^{3/2}\frac{E_{b}}{Y}\left(\frac{c}{c_{s}}\right)^{3}\right.\times\\ \nonumber
&&\left.\left(-\frac{2m}{r^{3}}+\frac{3\tilde{L}^{2}}{r^{4}}-\frac{12m\tilde{L}^{2}}{r^{5}}\right)^{3/2}\right)^{2/5}
\end{eqnarray}
 Like in solid-state materials, we can think of $\sigma/Y$ like a hardness of the magnetized clump of plasma. 
  In GC15 we argued that the mechanical
 binding energy $E_{b}$ in (\ref{eq11}), stored in the clump and keeping it together, should be 
 at least of the same order of that observed in HF QPOs, if the HF QPOs are produced 
 by the tidal disruption of the clump.
 The amplitude of HF QPOs is some percent the luminosity of the source, i.e. $L_{QPO}\sim 10^{35}-10^{36}$ erg/s in atoll sources. 
 Following the results in Section~\ref{sec3} this energy is emitted over a time scale of the order of $\sim 0.01$ s, 
 thus the energy of the QPO is $E_{QPO}\sim 10^{33}-10^{34}$ erg. However, 
 this observed energy is only some percent of the total energy emitted. 
 It is the energy modulated by Doppler boosting. 
 For a hot-spot with an overbrightness twice the background disk the modulated energy is only of the order of 1\% 
 \cite{2004ApJ...606.1098S}.
Thus, the total energy emitted would be $E_{b}\sim 10^{36}$ erg. Substituting this 
 $E_{b}$ in (\ref{eq11}) we get $\sigma/Y\sim 300$ for an atoll source with luminosity $L\sim10^{37}$ erg/s 
 (see also Sec.~IV and Sec.~VII B in GC15). 
 We can estimate the magnetic field of the clump of plasma. 
 Indeed, if the $E_{b}$ above is 
 the magnetic binding energy keeping the clump together, then $\sigma=300 Y=300 \rho c_{s}^{2}$
 is the magnetic pressure $P_{m}=B^{2}/{2 \mu_{0}}$, $B$ the magnetic field and 
 $\mu_{0}=4\pi \times 10^{-7}$ H/m is the magnetic permeability.       
 Equating $P_{m}$ to $\sigma$ (in Pascal) we derive a magnetic field permeating the 
 clump of $B\sim 5\times 10^{9}$ G. In the case of a Z-source, whose ratio was estimated in GC15 
 $\sigma/Y\sim 70$, inserting $\rho$ and $c_{s}$ for a Z-source with luminosity
 $L\sim 2\times 10^{38}$ erg/s we get $B\sim 10^{10}$ G, a larger value
 than atoll sources, as measured \cite{1999A&AT...18..447P}. 
 Note however that in Fig.~3 of Ref.~\cite{1999A&AT...18..447P}
 atoll sources are located in the region around $B\sim 5\times 10^{8}$ G,
 while Z-sources in that with $B\sim 5\times 10^{9}$ G. 
 The discrepancy between these values and those calculated here may be because we did a crude estimation here. 
 For example, we are using the vacuum magnetic permeability $\mu_{0}=4\pi\times 10^{-7}$ H/m,
 usually also used in plasmas. However, it may be different in the plasma we are dealing with.
On the other hand, tidal stretching simulations of the magnetic field 
 in a star \cite{2017MNRAS.469.4879B} show that the magnetic field of 
 the squeezed star strengths at least by a factor of 10. Thus, if HF QPOs 
 are related to the energy emitted by a magnetic clump of plasma stretched by tides, 
 the estimation of $B$ shown here could give a $B$ actually larger than that of the host LMXB.\\
Although this result is interesting, giving a $B$ consistent
 with that measured in NS LMXBs ($B\sim10^{8}-10^{13}$ G \cite{1999A&AT...18..447P,2015SSRv..191..293R}),
 we would stress that the issues in this section need close attention 
 in dedicated future works.    

\subsection{Amplitude of the detectable modulation}\label{sec4a}   

 Numerical simulations of a hot-spot orbiting around a Kerr black hole and emitting 
 photons show modulations detected as HF QPOs if the hot-spot has 
 some overbrighteness with respect to the disk \cite{2004ApJ...606.1098S}. 
 An overbrightness twice the 
 background disk can give HF QPOs with an amplitude of the order of 
 $\sim 1\%$ the luminosity of the hot-spot. The light curve of the orbiting hot-spot is modulated at the orbital period 
 because of Doppler boosting of the emitted photons, 
 such as relativistic beaming, and gravitational lensing \cite{2004ApJ...606.1098S}. 
 These relativistic effects magnify the intensity of the electromagnetic radiation emitted. 
 In the case of relativistic beaming,
 the magnification depends on the velocity of the hot-spot with respect to the observer 
 (see e.g. Ref.~\cite{2007Ap&SS.311..231K}) 
\begin{equation}\label{eq12}
I_{\nu (o)}=I_{\nu (e)} D^{p}
\end{equation}
 where $I_{\nu (o)}$ and $I_{\nu (e)}$ are the observed and emitted specific intensity $I_{\nu}$, $p=3+\alpha$  
 with $\alpha$ energy spectral index\footnote{In atoll NS LMXBs $\alpha\geq 1$ 
 (see e.g. Ref.~\cite{2005ApJ...626..298T})}, 
 $D$ is the Doppler factor 
\begin{equation}\label{eq13}
D=\frac{1}{\gamma\left(1-\beta\cos\left(\theta\right)\right)}
\end{equation} 
where $\gamma=1/\sqrt{(1-\beta^{2})}$ is the Lorentz factor and $\beta=v/c$, with $v$ orbital speed 
 of the clump and $c$ speed of light. Because we are investigating an interval of orbital 
 radii ranging $r\sim 6-13\ r_{g}$ it would be worth checking the relative Doppler boosting 
 at $6\ r_{g}$ and $13\ r_{g}$.
 The Lorentz factor $\gamma$ and the ratio $\beta$ at these two radii are 
 $(\beta, \gamma)_{13r_{g}}=(0.23745,1.02944)$ and $(\beta, \gamma)_{6r_{g}}=(0.35482,1.06959)$. 
For an edge-on view ($\theta=0$), inserting in (\ref{eq13}) these numbers 
 the relative increment of $D^{4}$ is by\footnote{For an inclination, e.g., $\theta= 50$ the relative magnification drops 
 to 24\%.} 67\%. Thus, this relative increment 
 affects by $0.67\ I_{\nu (o)}$ any intrinsic trend of $I_{\nu (o)}$ over $r\sim 6-13\ r_{g}$. 
 
 In Fig.~\ref{fig3} the energy that could be released
 and possible converted into radiation is in the interval of $2-8\times 10^{35}$ erg, for an 
 atoll source with a luminosity of $L_{atoll}\sim10^{37}$ erg/s. Over the time-scale the energy is released, 
 $\sim 0.01$ s, the background energy of the source then is $E_{atoll}\sim10^{35}$ erg. Therefore,
 we may have a clump of plasma a factor 8 brighter than the background radiation. 
 Following the results in Ref.~\cite{2004ApJ...606.1098S}, in which an overbrighteness of the 
 hot-spot by a factor of 2 turns modulations of $\sim1\%$, we may have modulations up to $\sim 4\%$, i.e. 
 of the order of $\sim 10^{33}-3\times 10^{34}$ erg. Thus, the amount  of orbital energy 
 released by the clump during tidal circularization of the orbit might  
 give modulations that could be detected at $\nu_{k}+\nu_{r}$   
 in the power spectrum. The mechanism to produce energy proposed here might 
 justify how the orbiting hot-spot would have the overbrightness claimed 
 in other works, in order to produce detectable HF QPOs 
 \cite{1999ApJ...524L..63S,2004ApJ...606.1098S,2014MNRAS.439.1933B}.

\begin{figure}[!t!]
\includegraphics[width=0.47\textwidth]{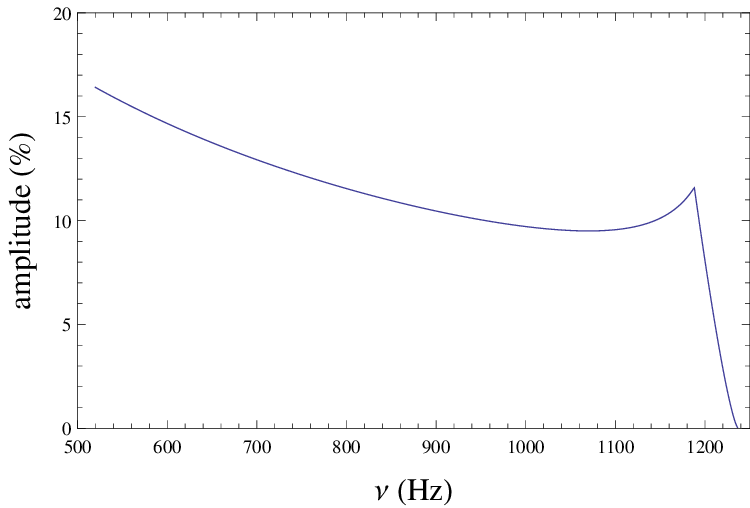}
\caption{Amplitude the beat $\nu_{k}+\nu_{r}$ would have in the observations
 after the energy release by tidal circularization of relativistic orbits (Fig.~\ref{fig3}). 
 The amplitude is in percent of the luminosity of an atoll NS LMXB ($\sim 10^{37}$ erg/s).  
 The amplitude is plotted as a function of the frequency of the beat $\nu_{k}+\nu_{r}$.
 Such behavior is typical of the amplitude of the upper HF QPO. For a comparison with the data 
 see Fig.~3 in Ref.~\cite{2006MNRAS.370.1140B}.}\label{fig5}
\end{figure}

 We divide the modulated fraction of energy by the time-scale 
 the tidal circularization of the orbit takes place, i.e. the time-scale over which 
 the energy is emitted, as a function of the orbital radius.
 Fig.~\ref{fig5} shows the amplitude the beat $\nu_{k}+\nu_{r}$ would have  
 in percent of the luminosity of the source $\sim 10^{37}$ erg/s.   
Both the value and the behavior in the figure are similar to the upper HF QPO amplitude  
 seen in the observations 
 (Fig.~3 in Ref.~\cite{2006MNRAS.370.1140B} (filled stars)), 
 where it is seen to decrease from $\sim 10-15\%$ to 1\% over the range of frequencies $\sim 500-1200$ Hz.

\section{Discussion}\label{sec5}

 Several models have been proposed in order to identify the central frequency of the twin-peak HF QPOs
 with those of the orbital motion around the compact object 
 \cite{1998ApJ...508..791M,1999ApJ...524L..63S,2004ApJ...606.1098S,2014MNRAS.439.1933B,2016MNRAS.457L..19T,
 1999ApJ...518L..95T,2003ApJ...584L..83M}.
 Some models link the keplerian frequency $\nu_{k}$ of the orbiting matter to the upper
 peak of the twin-peak HF QPOs, other link $\nu_{k}$ to the lower peak 
 \cite{1998ApJ...508..791M,1999ApJ...524L..63S,1999ApJ...518L..95T,2003ApJ...584L..83M}. 
 In Ref.~\cite{2009AIPC.1126..367G} numerical simulations show  
 that tidal disruption of clumps of matter 
 \cite{2009A&A...496..307K} produces power spectra much alike to the observed ones. 
 The power law and twin-peaks seen in the observations are reproduced. The upper peak corresponds to 
 $\nu_{k}+\nu_{r}$, the lower one to $\nu_{k}$. The light curve of an orbiting 
 clump/hot-spot is drawn by the timing law of its azimuth phase $\phi(t)$.  
 The photons emitted by the clump are cyclically Doppler boosted by relativistic effects    
 and when this happens is dictated by the timing law $\phi(t)$. Because in a curved space-time    
 for non-circular orbits $\nu_{r}\neq\nu_{k}$ the different orbital speed at periastron and 
 apastron passage introduces an oscillating term in $\phi(t)$ at the frequency $\nu_{r}$. 
 In a flat space-time $\phi(t)$ displays this oscillating term as well, 
 but in that case $\nu_{r}=\nu_{k}$ and in the power spectra of $\phi(t)$ only the peak 
 at $\nu_{k}$ is seen. In a curved space-time the beats $\nu_{k}\pm\nu_{r}$ and $\nu_{k}$ 
 are seen \cite{2013MNRAS.430L...1G}. 
 The beats at $\nu_{k}\pm\nu_{r}$ and $\nu_{k}$ are a characteristic of 
 the orbital motion as much as $\nu_{k}$ is in the case of a flat space-time. 
 Therefore, if orbital motion 
 in a curved-space time is producing the twin-peak HF QPOs, it is more natural to link 
 the upper HF QPO to $\nu_{k}+\nu_{r}$. This is also what numerical simulations 
 show \cite{2009AIPC.1126..367G,2013MNRAS.430L...1G}.
It is interesting noting that in the BH LMXB XTE J1550-564   
 was reported the evidence of a triplet of HF QPOs in a harmonic relationships, 92:184:276 Hz  
 \cite{2002ApJ...580.1030R}. The one at 92 Hz is the weakest. 
 Individual observations show only a HF QPO, but when averaged together to increase
 the signal to noise ratio the triplet show up. It is unlikely the same 
 HF QPO is going up and down in frequency since HF QPOs in BH LMXBs are observed at fixed 
 frequencies. Moreover, it would be a really unlikely occurrence the same peak
 showing up only at these three different orbital radii in integer 
 frequency ratios, 1:2:3.  
 The triplet would fit to the case in which the uppermost peak is 
 the beat $\nu_{k}+\nu_{r}$, while the other are $\nu_{k}$ and $\nu_{k}-\nu_{r}$ 
 (see also Ref.~\cite{2013MNRAS.430L...1G}). 
 The only orbital radius producing the triplet with 92:184:276 Hz 
 is $r_{p}\sim 7.3\ r_{g}$ for a Schwarzschild black hole with mass  
 $M_{BH}\sim 7.7\ M_{\odot}$. The mass 
 estimated from optical observations is $M_{BH}=9.10\pm 0.61\ M_{\odot}$ 
 \cite{2011ApJ...730...75O}. Therefore, the pairs of frequency ($\nu_{k}$, $\nu_{k}+\nu_{r}$), 
 given by numerical simulations \cite{2009AIPC.1126..367G}, is suitable for 
 interpreting the harmonic relationships of the HF QPOs seen in XTE J1550-564.\\
In Ref.~\cite{2014MNRAS.437.2554M} both the mass $M_{BH}$ and dimensionless 
 angular momentum $a$ of the BH LMXB GRO J1655-40 were measured by means of numerical fits, 
 linking $\nu_{k}$ to the upper peak 
 ($\sim 450$ Hz) while $\nu_{k}-\nu_{r}$ (periastron precession) to the lower one ($\sim 300$ Hz),
 as previously proposed by the model \cite{1999ApJ...524L..63S}.
It is not straightforward making a direct 
 comparison of the GRO J1655-40 mass measured in 
 Ref.~\cite{2014MNRAS.437.2554M}, 
 using the frequency pairs ($\nu_{k}-\nu_{r}$,
 $\nu_{k}$), to that using ($\nu_{k}, \nu_{k}+\nu_{r}$) as here suggested.  
 In Ref.~\cite{2014MNRAS.437.2554M} 
 relativistic frequencies in the Kerr metric were used to fit the data. Also,
 a third low frequency QPO ($\sim 18$ Hz) linked to the modulation at the nodal precession 
 frequency $\nu_{nod}$ was used in the fit. The precession of the plane of the orbit would 
 produce a modulation at $\nu_{nod}$, a general relativistic effect due to 
 frame dragging and known as Lense-Thirring precession \cite{1918PhyZ...19..156L}.
 In this manuscript we are using relativistic frequencies of low eccentricity orbits 
 in the Schwarzschild metric, since here we needed to use exact analytical expressions
 for both the energy $\tilde{E}$ and angular momentum $\tilde{L}$ for orbits 
 with generic eccentricity $e$ \cite{1994PhRvD..50.3816C}.
 Moreover, in the Schwarzschild metric the nodes of the orbit do not precess.
The mass of GRO J1655-40 from the fit in Ref.~\cite{2014MNRAS.437.2554M}  
 agrees with great accuracy to that from optical observations.
 The best-guess from optical light curves is $M_{BH}=5.4\pm 0.3\ M_{\odot}$
 \cite{2002MNRAS.331..351B}. The radius at which the three QPOs would be emitted
 in Ref.~\cite{2014MNRAS.437.2554M} is $r\sim 5.6\ r_{g}$, assuming that the
 low frequency QPO is the nodal frequency $\nu_{nod}$ 
 and not $2\nu_{nod}$ as originally proposed by the model \cite{1999ApJ...524L..63S}.   
 Using the frequency pairs ($\nu_{k}$, $\nu_{k}+\nu_{r}$)
 to produce twin-peak HF QPOs in a 3:2 ratio, 
 with the lower HF QPO $\sim300$ Hz and the upper $\sim450$ Hz 
 as in the observations, the mass of the 
 Schwarzschild black hole is $M_{BH}=4.7\ M_{\odot}$, and 
 the orbital radius where ($\nu_{k}$, $\nu_{k}+\nu_{r}$) are in
 3:2 ratio is\footnote{Note that in the Kerr metric this orbital radius would be 
 $\sim 7\ r_{g}$ for a Kerr black hole with $M_{BH}\sim 5.7\ M_{\odot}$, $a\sim 0.3$. 
 At this radius, the low frequency QPO ($\sim 18$ Hz) used in the fit in Ref.~\cite{2014MNRAS.437.2554M} 
 is equal to $2\nu_{nod}$.} 
 $r\sim 7.3\ r_{g}$.\\ 
 We emphasize that a precise measurement of the mass 
 of a compact object using the twin-peak HF QPOs is beyond the purpose of this 
 manuscript. It demands close attention and accurate methodology, like that 
 described in Ref.~\cite{2014MNRAS.437.2554M}. 

 In Ref.~\cite{2015ApJ...798L..29B} is reported an observational result 
 that could challenge the results presented in this manuscript,
 i.e. the upper HF QPO corresponding to $\nu_{k}+\nu_{r}$ (as numerical simulations 
 \cite{2009AIPC.1126..367G} and Figs.~\ref{fig4},~\ref{fig5} suggest). 
 The authors studied the behavior of the
 pulse amplitude in the accreting milliseconds x-ray pulsar SAX J1808.4-3658. It was noted,
 for the first time, that the pulse amplitude correlates with the  
 frequency (300-700 Hz) of the upper HF QPO detected. 
 It is shown that when the upper HF QPO frequency is below the spin
 frequency (401 Hz) of the pulsar,
 the pulse amplitude doubles. When the frequency of
 the upper HF QPO is above the spin frequency the pulse amplitude halves. 
 This shows evidences on a direct interaction between the 
 spinning magnetosphere of the neutron star and the physical mechanism producing 
 the upper HF QPO. It strongly suggests that the upper HF QPO
 originates from orbital motion of the plasma in the accretion disk.
 The possible keplerian nature of the upper HF QPO is highlighted.
 On the other hand, it is emphasized that the findings also suggest a more general 
 azimuthal nature of the upper HF QPO. It could be keplerian, precessional, or 
 an azimuthally propagating disk wave. If orbital motion 
 is producing the detected upper HF QPO, the findings in 
 Ref.~\cite{2015ApJ...798L..29B} would not discard an upper HF QPO 
 corresponding to the beat $\nu_{k}+\nu_{r}$, since this beat is a natural consequence 
 of orbital motion in the curved-space time around the spinning neutron star.\\ 
It is interesting noting that if the upper HF QPO ranging 300-700 Hz in SAX J1808.4-3658   
 is the beat $\nu_{k}+\nu_{r}$, it would correspond to a range of keplerian frequency
 $\nu_{k}\sim 200-400$ Hz, i.e. an upper limit equal to the 
 spin frequency of the magnetosphere (401 Hz). The maximum keplerian frequency
 then is seen at the corotational radius $r_{c}$, i.e. the orbital radius at which
 the keplerian frequency equals the spinning one. In Ref.~\cite{2008ApJ...675.1468H}
 has been suggested that SAX J1808.4-3658 is near spin equilibrium,
 i.e. $r_{m}\sim r_{c}$, where $r_{m}$ is the magnetosphere radius. Therefore, 
 a maximum upper HF QPO of $\sim 700$ Hz might mean a coherent oscillation 
 produced close to or at the magnetosphere radius. Either the 
 disk is truncated at the magnetosphere radius $r_{m}$ or inside 
 the magnetosphere no coherent oscillations form. 
Within this interpretation, 
 from the observations \cite{2015ApJ...798L..29B} we see that 
 as long as the upper HF QPO is produced
 closer and closer to $r_{m}$, the pulse amplitude of the neutron star decreases. 
 Following the arguments in Ref.~\cite{2015ApJ...798L..29B} on centrifugal inhibition,
 the interpretation of the upper HF QPO equal to the beat $\nu_{k}+\nu_{r}$ and, therefore, 
 $\nu_{k}\sim 200-400$ Hz may give suitable arguments.   
 When the plasma in the accretion disk orbits far away the magnetosphere, $r>r_{m}$,
 or $\nu_{k}<\nu_{s}$, the centrifugal force at the magnetosphere would inhibit this plasma accreting. 
 Therefore, for a clump of plasma orbiting in the disk and producing the upper HF QPO,
 some plasma of the clump would not be able to flow towards the 
 magnetic poles and would not affect the pulse amplitude. Instead,
 a clump of plasma orbiting closer to the 
 corotational radius, or close the magnetosphere, thus for keplerian frequencies approaching
 $\nu_{k}=401$ Hz and for $\nu_{k}+\nu_{r}$ above 400 Hz, it would be more
 likely that a fraction of the clump is accreted towards
 the poles, weakening the pulse amplitude  
 \cite{2015ApJ...798L..29B}. This interpretation, rather than an upper HF QPO 
 equal to $\nu_{k}$, might be more suitable for the excursions seen in the pulse amplitude of
 SAX J1808.4-3658. Such excursions cluster around a frequency of the upper HF QPO of
 $\sim 600-700$ Hz \cite{2015ApJ...798L..29B}, i.e. at $\nu_{k}\sim 330-400$ Hz, close 
 to the frequency at the corotational/magnetosphere radius (401 Hz), where some rest of
 the clump is more likely to flow to the magnetic poles, causing the pulse amplitude to flicker.

 Simultaneous twin-peak HF QPOs in SAX J1808.4-3658 are rarely seen. When 
 HF QPOs were discovered in this source \cite{2003Natur.424...44W},
 the twin-peaks were detected only in one observation. 
 A systematic study on the variability of SAX J1808.4-3658 has been presented   
 in Ref.~\cite{2015ApJ...806...90B}. Twin-peak HF QPOs were 
 detected only in three observations (out of many) with different central frequencies. 
 These three detections give clues on the evolution of the twin-peaks frequency.
 The separation in frequency of the
 peaks is almost consistent with a constant value ($\sim 180$ Hz)
 close to half the spin frequency of the pulsar, as 
 previously reported \cite{2003Natur.424...44W}. The highest 
 frequency of the upper HF QPO is $\sim 730$ Hz yet may be consistent with 
 the fact that the upper HF QPO corresponds to $\nu_{k}+\nu_{r}$ 
 and the highest upper HF QPO of $\sim 730$ Hz is produced at the 
 corotational/magnetosphere radius. On the other hand, a constant
 separation in frequency of twin-peaks is inconsistent with the pairs 
 ($\nu_{k}$, $\nu_{k}+\nu_{r}$), since the difference $\nu_{r}$ varies and 
 does not match the separation measured.
 However, a constant separation in frequency is a feature
 not seen in other atoll sources. 
 The separation usually varies by several tens of hertz 
 with varying central frequency of the peaks \cite{2006MNRAS.370.1140B}.
The lower HF QPOs in SAX J1808.4-3658 
 shows properties that make it to differ than the lower HF QPO in other atoll 
 sources. 
 In SAX J1808.4-3658 the upper HF QPOs is more prominent than the lower \cite{2015ApJ...806...90B}.
 When detected simultaneously, in other atoll sources the lower HF QPO shows a 
 larger amplitude \cite{2006MNRAS.370.1140B,2006MNRAS.371.1925M}. 
 The coherence $Q\sim 10$ of the lower HF QPO in SAX J1808.4-3658
 (of the same order of the upper one) \cite{2015ApJ...806...90B}
 is much lower than in other atoll NS LMXBs,
 where it can be of the order of $Q\sim 100$ \cite{2006MNRAS.370.1140B,2006MNRAS.371.1925M}. 
 Calculations in GC15 show that such high coherences may
 be typical of a keplerian modulation.\\ 
If the upper HF QPO in SAX J1808.4-3658 is the
 beat $\nu_{k}+\nu_{r}$ it might justify why its maximum frequency
 is $\sim 700$ Hz, since this frequency corresponds 
 to a keplerian frequency almost equal to the spinning one ($401$ Hz). 
 Therefore, coherent oscillations can form up to the corotational/magnetosphere 
 radius $r_{m}$, since the source is in spin equilibrium \cite{2008ApJ...675.1468H}.  
 Either the disk is truncated at the magnetosphere or inside no coherent modulations form.
 When the energy of such oscillations is released close to $r_{m}$ 
 the interaction with the magnetosphere might cause the excursions in pulse amplitude seen
 in SAX J1808.4-3658 \cite{2015ApJ...798L..29B}. The lower 
 HF QPO in SAX J1808.4-3658 might be a modulation different than keplerian 
 \cite{2003Natur.424...44W,2004ApJ...603L..89K}. It is rarely detected and
 shows different properties than the lower HF QPO detected in other atoll NS LMXBs.

\section{Conclusions}\label{sec6}

The power spectra of LMXBs are characterized by several peaks ranging from low to high
 frequencies. The highest frequencies detected often show up in pairs,  
 named twin-peak HF QPOs. They have central frequencies 
 typical of the orbital motion of matter  
 close to the compact object \cite{2016IJAA....6...82W}.\\ 
In atoll NS LMXBs the lower and
 upper HF QPOs show different patterns
 of their amplitude and coherence versus central frequency
 \cite{2001ApJ...561.1016M,2006MNRAS.370.1140B,2006MNRAS.371.1925M,2011ApJ...728....9B}. 
 The lower HF QPO shows an increase
 and then a decrease of its both amplitude and coherence. The amplitude of the 
 upper HF QPO keep decreasing with increasing central frequency of the peak. 
 The trend of its coherence remains of the order of $Q\sim 10$ over a large range of frequencies.
 Following numerical simulations \cite{2009AIPC.1126..367G}, in GC15 we have
 proposed that the lower twin-peak HF QPO could 
 arise from the energy released during tidal disruption of clumps 
 orbiting in the accretion disk.
 Here we have wondered whether the energy and coherence observed in 
 the upper HF QPO could originate because of the tidal 
 circularization of the clump's orbit.  
 The tidal force acting on an orbiting clump circularizes and
 shrinks the orbit and the clump emits the released orbital energy as 
 radiation \cite{2008A&A...487..527C}. 
 The modulation at $\nu_{k}+\nu_{r}$ caused by the eccentricity 
 of the orbit \cite{2013MNRAS.430L...1G} should originate because of the
 energy released in the phase of tidal circularization of the orbit. We have 
 estimated the energy that clumps of plasma orbiting in the accretion disk
 would release because of tidal circularization of their relativistic
 orbits.
 We note for the first time that such physical mechanism 
 might account for the \text{amplitude and coherence} of the upper HF QPO 
 observed in atoll NS LMXBs (Figs.~2,~3 of Ref.~\cite{2006MNRAS.370.1140B}). 
 Numerical simulations \cite{2009AIPC.1126..367G,2013MNRAS.430L...1G}, the 
 results presented here (Figs.~\ref{fig4},~\ref{fig5}) and the discussion on 
  SAX J1808.4-3658 suggest that the upper 
 HF QPO most probably corresponds to the beat $\nu_{k}+\nu_{r}$.
 
The physical mechanism to release energy proposed here, together with the modulation mechanism
 in Refs.~\cite{2004ApJ...606.1098S,2009AIPC.1126..367G,2013MNRAS.430L...1G,2014MNRAS.439.1933B},
 might offer an explanation on why the upper HF QPO 
 would originate. This work might be the first time we are recognizing the 
 tidal circularization of relativistic orbits occurring around a neutron star.

\begin{acknowledgments}
I would like to thank Rodolfo Casana, Manoel M. Ferreira Jr., Adalto R. Gomes and Alessandro Patruno 
for discussions on the topic. I thank the anonymous referees 
 for valuable comments that helped to improve the manuscript. 
This work was fully supported by the program PNPD/CAPES-Brazil.
\end{acknowledgments}

\bibliography{biblio.bib}

\end{document}

%% file: ads.tex
\def\araa{ARA\&A}%
\def\apj{ApJ}%
\def\apjl{ApJ}%
\def\apss{Ap\&SS}%
\def\aap{A\&A}%
\def\icarus{Icarus}%
\def\mnras{MNRAS}%
\def\prd{Phys.~Rev.~D}%
\def\ssr{Space~Sci.~Rev.}%
\def\nat{Nature}%